\documentclass[10pt, twocolumn]{IEEEtran}
\usepackage[utf8]{inputenc}
\usepackage{cite}
\usepackage{todonotes}
\usepackage{multirow}
\usepackage{booktabs}
\usepackage{amsmath}
\usepackage{amssymb}
\usepackage{url}
\usepackage{textcomp}
\usepackage{caption}
\usepackage{subcaption}

\begin{document}
\bibliographystyle{IEEEtran}
\title{Optimal Steerable mmWave Mesh Backhaul Reconfiguration}
\author{
\IEEEauthorblockN{Ricardo Santos$^{*}$, Hakim Ghazzai$^{\#}$, and Andreas Kassler$^{*}$\\
\IEEEauthorblockA{\small $^{*}$Karlstad University, Karlstad, Sweden\\ Email: \{ricardo.santos, andreas.kassler\}@kau.se\\
$^{\#}$Stevens Institute of Technology, Hoboken, NJ, USA,\\ Email: hghazzai@stevens.edu}
{\thanks {\vspace{-0.4cm}\hrule
\vspace{0.1cm} \indent 2018 IEEE. Personal use of this material is permitted. Permission from IEEE must be obtained for all other uses, in any current or future media, including reprinting/republishing this material for advertising or promotional purposes, creating new collective works, for resale or redistribution to servers or lists, or reuse of any copyrighted component of this work in other works.}}
}
\vspace{-0.4cm}}


\maketitle
\thispagestyle{empty}
\pagenumbering{arabic}

\begin{abstract}
Future 5G mobile networks will require increased backhaul (BH) capacity to connect a massive amount of high capacity small cells (SCs) to the network. Because having an optical connection to each SC might be infeasible, mmWave-based (e.g. 60 GHz) BH links are an interesting alternative due to their large available bandwidth.
To cope with the increased path loss, mmWave links require directional antennas that should be able to direct their beams to different neighbors, to dynamically change the BH topology, in case new nodes are powered on/off or the traffic demand has changed. Such BH adaptation needs to be orchestrated to minimize the impact on existing traffic.
This paper develops a Software-defined networking-based framework that guides the optimal reconfiguration of mesh BH networks composed by mmWave links, where antennas need to be mechanically aligned.
By modelling the problem as a Mixed Integer Linear Program (MILP), its solution returns the optimal ordering of events necessary to transition between two BH network configurations.
The model creates backup paths whenever it is possible, while minimizing the packet loss of ongoing flows.
A numerical evaluation with different topologies and traffic demands shows that increasing the number of BH interfaces per SC from 2 to 4 can decrease the total loss by more than 50\%.
Moreover, when increasing the total reconfiguration time, additional backup paths can be created, consequently reducing the reconfiguration impact on existing traffic.

\begin{IEEEkeywords}
5G, mesh backhaul, mmWave, optimization.
\end{IEEEkeywords}

\end{abstract}

\section{Introduction}
\label{sec:intro}
Current wireless technologies cannot easily cope with the increase of mobile users and respective data rate requirements. This motivates research for new radio (NR) technologies and backhaul (BH) network architectures for the next generation mobile networks, i.e. 5G\cite{chen2014requirements}. To increase the overall capacity, 5G features ultra-wide bandwidth links operating in very high frequency bands (6-60 GHz), including millimeter-wave (mmWave) spectrum, both for access and BH. The high path loss, combined with the unique propagation characteristics of mmWave links, require a massive amount of small cells (SCs) to be deployed \cite{7999294}. As it is economically unfeasible to connect all the SCs through optical cables, many wireless BH links may use mmWave frequencies, leading to multi-hop mesh topologies \cite{huang2017efficient} that connect the SC nodes to the umbrella macro cell, forming heterogeneous networks (HetNets).

Because of the high path loss, mmWave links require highly directional antennas that need to be aligned properly. For example, when using a 7$^\circ$ beam width, a misalignment of 18$^\circ$ reduces the link quality by 18 dB, compared to a perfect alignment \cite{7218630}, which may result in link breaks or capacity degradation. Therefore, when scaling the usage of mmWave links to an entire BH infrastructure, an efficient beam alignment becomes a crucial topic to address. Aperture antennas have been designed for planar form factors, which are compact, have high gain, and support wideband mmWave applications \cite{7312444}. Additionally, to support the alignment with a wide angular coverage, different techniques can be applied to enable the antenna radiation steering, ranging from optic reflection/refraction techniques to complex digital schematics or mechanical rotating components. Despite the possibility of failure due to the fatigue of moving parts in mechanical steerable antennas, they provide a flexible steering range, while maintaining the antenna gain \cite{uchendu2016survey}. The antennas can then be mounted on rotating devices that can be built using simple electronic components.

In addition to the traffic management complexity due to the high number of BH nodes and links, energy-related operational costs can increase and an energy-efficient network reconfiguration becomes crucial \cite{sakaguchi2017mmwave}. Examples include dynamically powering on/off network nodes/interfaces \cite{tran2016practical} or an energy-efficient joint user association, resource allocation and BH routing strategy \cite{mesodiakaki2017joint}. These reconfigurations can be temporary and change according to adaptive parameters (e.g. traffic load, overall system power consumption), or can be guided by policies defined by network operators. Yet, the reconfiguration process can be complex and involve different steps.  For example, if a new SC is powered on to serve more users in a given area, new links need to be established to cope with the topology change. Similarly, when BH nodes are powered down for energy conservation, their active flows must be rerouted. Moreover, all these requirements must be applied to the entire BH, which significantly complicates the reconfiguration. Existing work is mostly focused on effective flow migration techniques to avoid network disruption \cite{danielis2017dynamic}, yet an orchestrated reconfiguration of a mmWave BH is not properly addressed.

This paper presents a SC mesh BH reconfiguration mechanism designed for mechanical steerable antennas. We develop a mathematical optimization model based on a Mixed Integer Linear Program (MILP) that calculates the optimal sequence of flow re-routing and BH topology reconfiguration operations to allow the transition of the mesh BH from a suboptimal state to the desired one. The model minimizes the impact of the reconfiguration on existing BH traffic by calculating an optimized sequence of antenna re-alignment, link establishments and flow re-routing operations. The developed framework can be implemented through a Software Defined Networking (SDN) based architecture, where the SDN controller orchestrates the BH reconfiguration using the proposed model. We evaluate the model with different topologies, number of transceivers, and maximum allowed reconfiguration times. Our numerical results indicate that increasing the number of BH transceivers from 2 to 4 leads to more backup path creation opportunities, contributing to a decrease of the overall packet loss by more than 50\%. Also, by increasing the allowed reconfiguration timespan, further backup links can be formed, while the final configuration links are being established, through the rotation of their antennas.

This paper is organized as follows. Section~\ref{sec:model-sdn} describes the system model and the proposed SDN-based architecture, which motivates the problem formulation in Section~\ref{sec:problem}. Experimental results are presented in Section~\ref{sec:results}, concluding with final remarks and future work in Section~\ref{sec:conclusions}.

\section{System Model and SDN-based Architecture}
\label{sec:model-sdn}
In this section, we introduce the system model and present an overview of our SDN-based mmWave BH architecture.

\subsection{System Model Considerations}
\label{subsec:system-model}

\begin{figure}
\vspace{+0.3cm}
\centering \includegraphics[width=0.88\columnwidth]{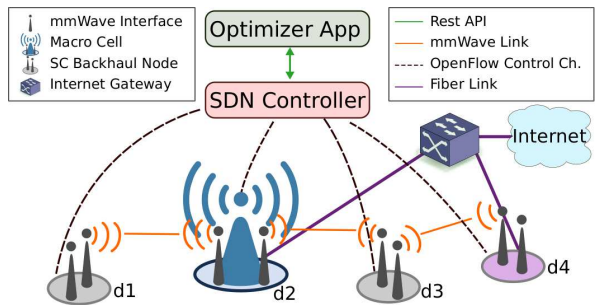}
\caption{System architecture of a SDN-based HetNet.}
\label{fig:architecture}
\vspace{-0.2cm}
\end{figure}

The main goal of this work is to guide the reconfiguration of a mmWave based SC mesh BH network, which operates under the umbrella of an attached macro cell, composed by a set $\mathcal{D}$ of mmWave SC nodes (Figure~\ref{fig:architecture}). Every BH node $d$, where $d=1 \dots D \in \mathcal{D}$, is located at $pos_d=[x_d, y_d]$, where $x_d$ and $y_d$ are its respective two dimensional space geographical coordinates. We assume that the nodes are located within close vicinity and, for simplicity,  have the same altitude. Every node has a set of network interfaces $\mathcal{N}$ with $N$ elements, each establishing BH links using a mmWave transceiver. Each interface has an antenna, which is mounted over a mechanical rotational platform that can horizontally rotate over 360\textdegree, with 0\textdegree ~being oriented to the absolute North and vertically between -45\textdegree ~and 45\textdegree. We assume that all antennas can rotate simultaneously with the same speed. We also focus on the network interfaces alignment on the horizontal axis, assuming that the vertical alignment can be neglected.

We assume that the BH is initially configured to serve a given user demand, defined by the user equipments (UEs) associated to the SCs (we identify this initial configuration by snapshot A). If the demands change, e.g. new users arrive and/or depart, the mesh BH may need to change to serve the new requirements. This reconfiguration time can be long due to the antennas' rotation, the update of forwarding rules, or due to powering on/off BH nodes. All these configuration steps may interrupt ongoing network traffic, if not orchestrated correctly among the BH. With that, we aim to optimally solve this transitioning problem from snapshot A to the new configuration (identified as snapshot B), with minimal disruption on active UE traffic. In this work, we are essentially interested in the ordered reconfiguration of the antennas' alignment, since the most time critical aspect is the rotation of the mechanical platforms, which may take several seconds. In fact, our previous measurements using an IEEE 802.11ad testbed have shown that re-establishing flow connectivity using a SDN controller can be achieved within 6 ms, once the antennas are aligned and the links are formed \cite{8269214}. During the reconfiguration, if time and resources allow, we aim to explicitly create backup paths, which can be used to temporarily route traffic, while the original link is down due to the antenna rotation.  

To form the mmWave BH mesh, links need to be established between two nodes and we consider that both interfaces have to be within line-of-sight and aligned with each other. We establish a matrix $\boldsymbol{V}$ with $D \times D$ elements, containing the angles necessary to align all the nodes' interfaces. Such matrix can be computed once, during an initial calibration, based on the nodes' geographical positions. A binary matrix $\boldsymbol{\delta}$ with size $D \times D$ lists all the possible connections between the nodes in $\mathcal{D}$. We assume that each link is characterized by an average throughput $R_{dd'}$, given by the average statistics of the channel, based on the path loss due to propagation and atmospheric conditions\cite{mesodiakaki2016energy}. We ignore effects of fast-fading, as we focus on a long time period of the network operation, compared to the channel coherence time.

We assume that a subset $\mathcal{I}\subseteq\mathcal{D}$ of the SC nodes is connected through a fiber link to the core network. Every mesh node $d$ serves a traffic demand $\rho_d$ (measured in Mbps), which is given by the aggregated demands of all the UEs served by node $d$. In this model, we only consider downlink traffic, as we assume that upstream traffic does not have high bandwidth requirements and can share the same downlink paths. Therefore, we model the paths by creating unidirectional links from the nodes in $\mathcal{I}$ to the rest of the topology.

\subsection{SDN-based mmWave Mesh Backhaul}
\label{subsec:SDN}
With the increase of the number of nodes and links in the BH network, a full reconfiguration involves several link and BH routing reconfigurations, which may be difficult to implement using distributed protocols, because they typically require a large time to converge. This motivates the usage of centralized BH management solutions. SDN allows the separation of the control plane from the data plane and the shifting of the network management to a logically centralized SDN controller, which has a global view of the network. Typically, control plane messages are exchanged using the OpenFlow (OF) protocol \cite{mckeown2008openflow}, that manages the forwarding rules in network nodes. Power management and wireless link configurations can then be handled by specific device management protocols (e.g. NETCONF \cite{enns2011network}), or through extensions to the original OF specification \cite{santos2017small}. This enables new ways for optimizing BH network reconfigurations, as the link configurations, forwarding paths, and powering on/off SCs can be centralized and orchestrated by a single entity \cite{SDNPOC2017}.

We assume that the SC mesh BH network is based on a SDN architecture, having the SDN controller responsible for the management of the mesh nodes' control plane, configuring their forwarding rules and respective mmWave links. We consider the usage of a SDN out-of-band control channel \cite{santos2016sdn}, with negligible latency between the controller and the mesh nodes. The SDN controller provides an application programming interface (API) that allows the specification of how the network should be configured, according to the model's output. After the reconfiguration steps are calculated, the controller pushes the new flow rules and requests the antenna rotations and link configurations to the BH nodes.

\section{Problem Formulation}
\label{sec:problem}
Given two network configuration snapshots (A and B), the main goal is to optimally orchestrate the transition from A to B, within a given maximum reconfiguration time. Since there are links and flows established in A, we want to minimize the total network packet loss during the reconfiguration.

\subsection{Time Horizon Discretization}
\label{subsec:time}

We assume that the time is divided into $K$ slots, where $K$ indicates the length of the time horizon, which represents the maximum duration of the transition phase allowed by the network operator. We denote the length of each time slot $k$, where $k=1,\dots, K$, by $\tau$. The first and final time slots, i.e. $k=1$ and $k=K$, correspond to the initial and last network configuration snapshots (A and B), respectively. The number of time slots $K$ must be large enough to accommodate all the necessary reconfiguration steps. Since our system operates by rotating network interfaces, we quantize their movements by setting $\tau$ to the time required to horizontally rotate an interface by $\theta$ degrees (we assume the same values for all interfaces). Therefore, the maximum configuration time for $K$ is given by $K\tau$, e.g. for $K=20$ and $\tau=0.5s$, the total reconfiguration time would be 10 seconds. The minimum $K$ for a given topology can be defined by the maximum number of horizontal angular movements required by a network interface, in multiples of $\theta$. In a worst case, this is equivalent to a 180\textdegree ~rotation (e.g. $K=19$, if $\theta=10$, assuming no movement at $k=1$).

\subsection{Decision Variables and Constraints}
\label{subsec:variables}
To ensure a safe operation of the network, we need to guarantee that all the links given by configuration B are established at the end of the reconfiguration. A binary variable $x^k_{dnd'n'}$ indicates whether an unidirectional link is established from node $d$ antenna $n$ to node $d'$ antenna $n'$ or not as follows:
 \begin{equation}
\label{eq:x-values}
   x^k_{dnd'n'} =\left\{
   \begin{array}{ll}
   1, & \hbox{if interface $n$ of node $d$ is connected} \\
   		& \hbox{to interface $n'$ of node $d'$ during the} \\
	 	& \hbox{$k^{\text{th}}$ time period},\\
   0, & \hbox{otherwise.}
                   \end{array}
                 \right.
\end{equation}

A link can only be formed between two interfaces if it possible to interconnect the respective nodes. Hence, the following constraint is added:
 \begin{equation}
\label{eq:deltas}
x^k_{dnd'n'} \leq \delta_{dd'}, \forall d, n, d', n', k.
\end{equation}

For unidirectional links and only allowing one interface to establish a single link with other interfaces, we add the following constraints:
 \begin{equation}
\label{eq:uni-direction}
\sum_{d' \in \mathcal{D}} \sum_{n' \in \mathcal{N}} x^k_{dnd'n'} + x^k_{d'n'dn} \leq 1, \forall d, n, k.
\end{equation}

To quantify the ingress rate entering the nodes connected to the core network in $\mathcal{I}$, we introduce a continuous decision variable $input^k_d$. We limit the ingress rate on all the nodes $d \in \mathcal{I}$ to not be negative. In addition, the total ingress rate can not exceed the total demand of all the nodes in $\mathcal{D}$:

\begin{subequations}
\label{eq:input-rate}
\begin{align}
& 0 \leq input^k_d, \forall d, k,\\
& \sum_{d \in \mathcal{I}} input^k_d \leq \sum_{d' \in \mathcal{D}} \rho_{d'}, \forall k.
\end{align}
\end{subequations} 

Secondly, we need to map the amount of traffic that is sent through every link, on every time slot $k$, and limit it to the maximum link capacity, defined in $R_{dd'}$. Since we use unidirectional links, we create a variable $r^k_{dnd'n'} \in \mathbb{R}^+ \geq 0$ that represents the data rate from interface $n$ of node $d$ to interface $n'$ of node $d'$, at the $k^{\text{th}}$ time slot:
\begin{equation}
\label{eq:link-rate}
0 \leq r^k_{dnd'n'} \leq R_{dd'}, \forall d, n, d', n', k.
\end{equation}

Ideally, this value should be big enough to fulfill the demands $\rho_d$ for every node, but since the maximum capacity or the absence of a link cannot always make that condition true, we use a variable $l^k_d \in \mathbb{R}^+$ to calculate the loss rate of every node $d$, on a given time slot $k$:
\begin{equation}
\label{eq:loss-constraint}
0 \leq l^k_d \leq \rho_d, \forall d, k.
\end{equation}

Additional constraints are imposed to limit the total acceptable packet loss rate on  different time slots (e.g. the loss rate should not exceed 45\% at any time), i.e.:
\begin{equation}
\label{eq:loss-threshold}
\sum_{d \in \mathcal{D}} l^k_d \leq v^k \left[ \sum_{d \in \mathcal{D}} \rho_d \right], \forall k,
\end{equation}

\noindent where $v^k \in [0, 1]$ controls the tolerated loss at time slot $k$, with respect to the demand.

The flow conservation constraints ensure, for every node, that the total input rate and packet loss are equal to the total output rate and its traffic demand. Also, data can only flow between established links. To the nodes connected to the core network $d \in \mathcal{I}$, the data from $input^k_d$ must also be added:
\begin{subequations}
\label{eq:conservation}
\begin{align}
& \sum_{d' \in \mathcal{D}}  \sum_{n' \in \mathcal{N}} \sum_{n \in \mathcal{N}} x^k_{d'n'dn} r^k_{d'n'dn} + l^k_{d} \notag\\
&= \sum_{d' \in \mathcal{D}} \sum_{n' \in \mathcal{N}} \sum_{n \in \mathcal{N}} x^k_{dnd'n'} r^k_{dnd'n'} + \rho_{d}, \;\forall d \in \mathcal{D} \setminus \mathcal{I}, \forall k,\\
& input^k_d + \sum_{d' \in \mathcal{D}}  \sum_{n' \in \mathcal{N}} \sum_{n \in \mathcal{N}} x^k_{d'n'dn} r^k_{d'n'dn} + l^k_{d} \notag\\
&=  \sum_{d' \in \mathcal{D}} \sum_{n' \in \mathcal{N}} \sum_{n \in \mathcal{N}} x^k_{dnd'n'} r^k_{dnd'n'} + \rho_{d}, \;\forall d \in \mathcal{I}, \forall k.
\end{align}
\end{subequations} 

We define $a^k_{dn} \in \mathbb{R}$ to represent the current orientation of network interface $n$ of node $d$, during time slot $k$. The rotation of an interface can be incremented by $\theta$ with positive or negative values. Two binary variables, $\psi^k_{dn}$ and $\omega^k_{dn}$, indicate the movement of an interface during each time slot, clock and counter-clockwise, respectively:
\begin{equation}
\label{eq:mc-values}
\psi^k_{dn} = \left\{
\begin{array}{ll}
1, & \hbox{if interface $n$ of node $d$ is rotating} \\
	& \hbox{clockwise during time period $k$},\\
0	& \hbox{otherwise.}
\end{array}
\right.
\end{equation} 

\begin{equation}
\label{eq:mcc-values}
\omega^k_{dn} = \left\{
\begin{array}{ll}
1, & \hbox{if interface $n$ of node $d$ is rotating} \\
	& \hbox{counter-clockwise during time period $k$},\\
0	& \hbox{otherwise.}
\end{array}
\right.
\end{equation} 

We guarantee that an interface can only move in one direction (or none) at the same time, by only allowing one of those two decision variables to be true:
\begin{equation}
\label{eq:one-direction}
\psi^k_{dn} + \omega^k_{dn} \leq 1, \forall d, n, k.
\end{equation} 

The orientation of a network interface at time $k$ is given by its previous value (at $k-1$) and its current movement state:
\begin{equation}
\label{eq:movement}
\begin{split}
a^{k+1}_{dn} = (\psi^k_{dn} - \omega^k_{dn}) \theta + a^k_{dn},\\
\forall d, \forall n, k=1,\cdots, K-1.
\end{split}
\end{equation}

To be able to establish a link between two interfaces, we must ensure that both are aligned with each other (although it is also possible that they are aligned, but no link is created). We use the values $V_{dd'}$, elements of $\boldsymbol{V}$, to create those constraints. Since $V_{dd'} \in [0, 360[$ and the values of $a_{dn}^k$ can have values in the whole space $\mathbb{R}$ due to the increments of \eqref{eq:movement}, we introduce a decision variable $\beta^k_{d,n} \in \mathbb{N}$, used to assert the interface positions to the values in $V_{dd'}$, when they exceed their bounds:
\begin{align}
\label{eq:x-aligned}
\textbf{if } & a^k_{dn} = V_{dd'} + \beta^k_{dn} 360 \textbf{ and } a^k_{d'n'} = V_{d'd} + \beta^k_{d'n'} 360 \notag\\
\textbf{then, } & x^k_{dnd'n'} + x^k_{d'n'dn} \leq 1,\notag\\
\textbf{else } & x^k_{dnd'n'} + x^k_{d'n'dn} = 0.
\end{align}

This allows the establishment of a link immediately after both interfaces are aligned. We neglect the link configuration delay between the two interfaces, as we consider them significantly lower than the mechanical alignment times.

The set of links established in the initial configuration snapshot is given by $\mathcal{X}_{init}$. Likewise, the target link configuration that must be achieved by the end of the model execution is given by $\mathcal{X}_{end}$.  Therefore, we add the following initialization constraints that guarantee that only the links in $\mathcal{X}_{init}$ and $\mathcal{X}_{end}$ are formed, during $k=1$ and $k=K$, respectively:
 \begin{equation}
\label{eq:init-x}
x^1_{dnd'n'} =
\begin{cases}
1 & \quad \text{if link } (dnd'n') \text{ is part of } \mathcal{X}_{init}, \\
0 & \quad \text{otherwise}.
\end{cases}
\end{equation}

 \begin{equation}
\label{eq:end-x}
x^K_{dnd'n'} =
\begin{cases}
1 & \quad \text{if link } (dnd'n') \text{ is part of } \mathcal{X}_{end}, \\
0 & \quad \text{otherwise}.
\end{cases}
\end{equation}

Finally, as an input to the representation of the initial state, a matrix with size $D \times N$, denoted by $\boldsymbol{A^0}$, provides the angle alignment of every network interface $n$ of node $d$. We then use its elements $\mathcal{A}_{dn}^0$ to guarantee the initial alignment of every interface on the first time slot $k=1$:
\begin{equation}
\label{eq:init-angle}
a^1_{dn} = \mathcal{A}^0_{dn}, \forall d, n.
\end{equation}

\subsection{Objective Function}
\label{subsec:objective}
The main goal is to minimize the total packet loss rate, which is experienced during the reconfiguration from the state $X_{init}$ until the $X_{end}$ state at the final time slot $K$. To guide the network towards different solution approaches, we use a weight function $m_k = f(k)$ to multiply the total packet loss on each time slot. For instance, to force a faster reconfiguration, the weight function can assign higher weights to higher $k$ values, leading to a more severe loss weighting at later time slots. With that, the objective function, denoted by $\mathcal{F}$, can be described as:
\begin{equation}
\label{eq:objective}
\mathcal{F} = \sum_{k=1}^K \sum_{d=1}^\mathcal{D}  m_k l_{d}^{k}.
\end{equation}

\subsection{Linearization}
\label{subsec:linearization}
Since not all formulated constraints are linear (e.g. \eqref{eq:conservation} contains the product of two decision variables), we linearize the respective constraints to be able to formulate a MILP.

To linearize the product between $x^k_{dnd'n'}$ and $r^k_{dnd'n'}$ in \eqref{eq:conservation}, we add a new decision variable $z^k_{dnd'n'} \in \mathbb{R}^+$ and the following linear constraints:
\begin{subequations}
\label{eq:z-linear}
\begin{align}
& z^k_{dnd'n'} \leq x^k_{dnd'n'} R_{d'd}, \;\forall d,n,d',n',k, \\
& z^k_{dnd'n'} \leq r^k_{dnd'n'}, \;\forall d,n,d',n',k, \\
& z^k_{dnd'n'} \geq r^k_{dnd'n'} - (1 - x^k_{dnd'n'}) R_{d'd}, \;\forall d,n,d',n',k, \\
& z^k_{dnd'n'} \geq 0, \;\forall d,n,d',n',k.
\end{align}
\end{subequations} 

\noindent This allow us to re-write both equations \eqref{eq:conservation} as follows:
\begin{subequations}
\label{eq:conservation-z}
\begin{align}
&\sum_{d' \in \mathcal{D}}  \sum_{n' \in \mathcal{N}} \sum_{n \in \mathcal{N}} z^k_{d'n'dn} + l^k_{d} \notag\\
&= \sum_{d' \in \mathcal{D}} \sum_{n' \in \mathcal{N}} \sum_{n \in \mathcal{N}} z^k_{dnd'n'} + \rho_{d}, \;\forall d \in \mathcal{D} \setminus \mathcal{I}, \forall k,\\
&input^k_d + \sum_{d' \in \mathcal{D}}  \sum_{n' \in \mathcal{N}} \sum_{n \in \mathcal{N}} z^k_{d'n'dn} + l^k_{d} \notag\\
&=  \sum_{d' \in \mathcal{D}} \sum_{n' \in \mathcal{N}} \sum_{n \in \mathcal{N}} z^k_{dnd'n'} + \rho_{d}, \;\forall d \in \mathcal{I}, \forall k.
\end{align}
\end{subequations} 

Equation \eqref{eq:x-aligned} can be linearized by introducing a binary variable $p^k_{dnd'}$ that is set to 1 when interface $n$ of node $d$ is aligned with node $d'$:
\begin{equation}
\label{eq:p-def}
p^k_{dnd'} = \left\{
\begin{array}{ll}
1, & \hbox{if $a^k_{dn} = V_{dd'} + \beta^k_{dn} 360$}, \\
0	& \hbox{otherwise.}
\end{array}
\right.
\end{equation}

We then add the following linear constraints using the big-~M technique, restricting the values of $x^k_{dnd'n'}$ to only be true when the corresponding links are aligned, i.e. when $p=1$:
\begin{subequations}
\label{eq:p-linear}
\begin{align}
& a^k_{dn} - (V_{dd'} + \beta^k_{dn} 360) \leq M(1-p^k_{dnd'}), \;\forall d,n,d',k, \\
& a^k_{dn} - (V_{dd'} + \beta^k_{dn} 360) \geq -M(1-p^k_{dnd'}), \;\forall d,n,d',k, \\
& x^k_{dnd'n'} + x^k_{d'n'dn} \leq p^k_{dnd'}, \;\forall d,n,d',n',k,
\end{align}
\end{subequations} 
\noindent where $M$ is a big positive number.

\subsection{Mixed Integer Linear Programming Optimization Model}
\label{subsec:milp}
In this section, we formulate a generic optimization problem $(P)$ that minimizes the total packet loss for a given topology, while transitioning from the initial to the final configuration states. As output, our problem returns the BH link configuration and mmWave interface movement information at each time slot $k$, where $k=1,\dots,K$:
\begin{align}
\text{(P): }\;&\underset{x,a,z,p, \atop \beta,\psi,\omega}{\text{minimize}} \quad\quad \sum_{k=1}^K \sum_{d=1}^D m_k l_{d}^{k}\label{obj}\\   
&\text{subject to}\quad\quad \eqref{eq:x-values}, \eqref{eq:deltas}, \eqref{eq:uni-direction}, \eqref{eq:input-rate},\eqref{eq:loss-constraint}, \eqref{eq:loss-threshold},\eqref{eq:mc-values}, \eqref{eq:mcc-values}, \notag\\ \notag 
& \eqref{eq:one-direction}, \eqref{eq:movement}, \eqref{eq:init-x}, \eqref{eq:end-x}, \eqref{eq:init-angle}, \eqref{eq:z-linear}, \eqref{eq:conservation-z}, \eqref{eq:p-linear}. \notag \\ \notag
\end{align}

Problem $(P)$ can be solved optimally using e.g. off-the-shelf tools such as YALMIP/Gurobi \cite{lofberg2004yalmip}, which use branch and bound techniques to calculate optimal solutions. It is known that these problems are computationally expensive. Yet, this complexity does not pose a critical issue since the problem will be solved once before every transition phase which is, in practice, not recurrent. In addition, (P) can serve as a benchmark algorithm that can be used to compare fast solution heuristics. However, the development of such heuristics is outside the scope of this paper and left for a future extension.

\section{Numerical Results}
\label{sec:results}
We evaluate our model using three topologies, each with a different number of mesh nodes $D$ and by varying the number of network interfaces $N$, as presented in Table~\ref{table:topologies}. Each topology has a single node connected to the core network $d \in \mathcal{I}$ that forwards all the traffic to the rest of the BH nodes.

\begin{figure*}%
\vspace{+0.1cm}
\centering
\begin{subfigure}{.5\columnwidth}
\includegraphics[width=\columnwidth]{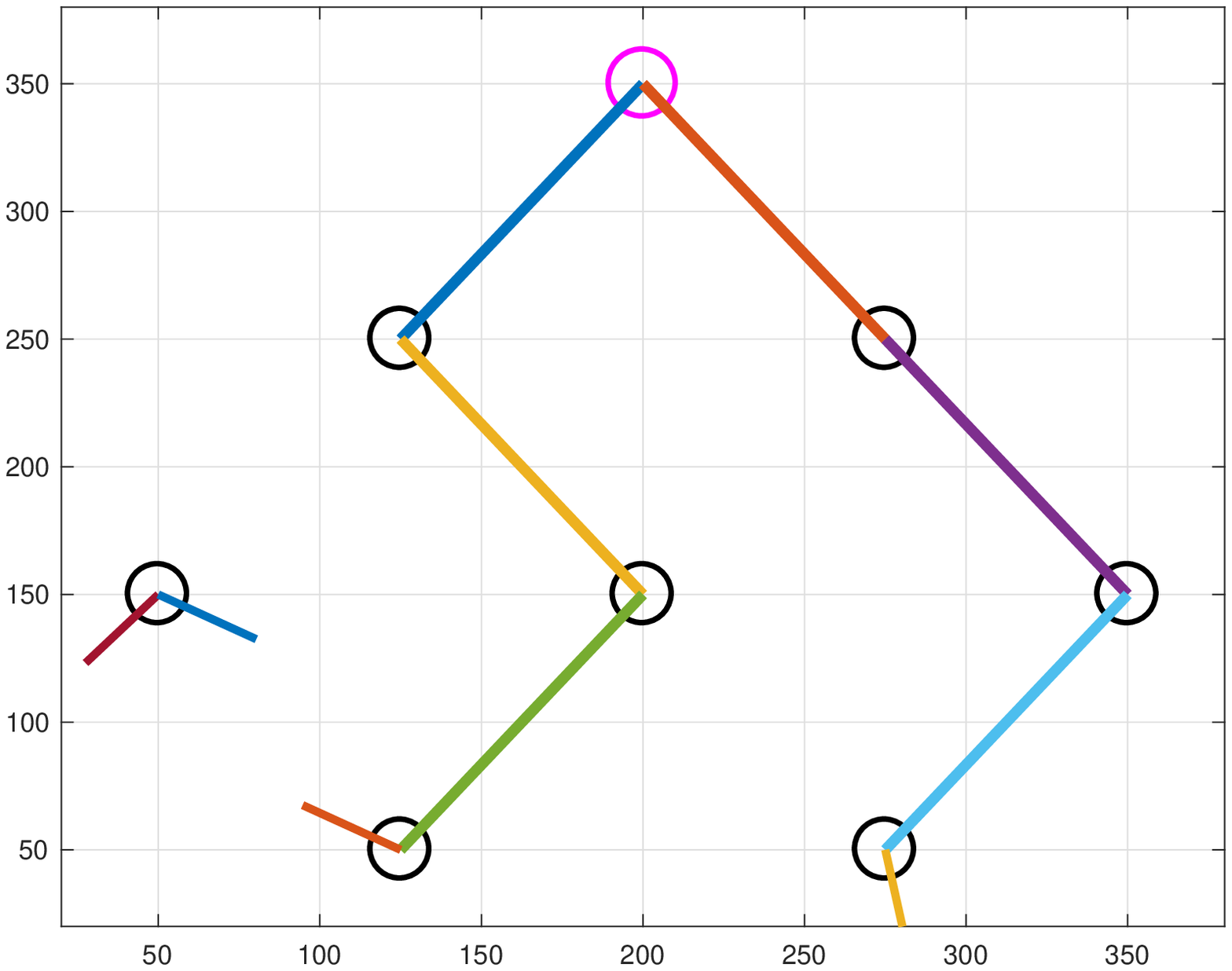}%
\caption{Initial state ($k=1$)}%
\label{subfig-simple1}%
\end{subfigure}\hfill%
\begin{subfigure}{.5\columnwidth}
\includegraphics[width=\columnwidth]{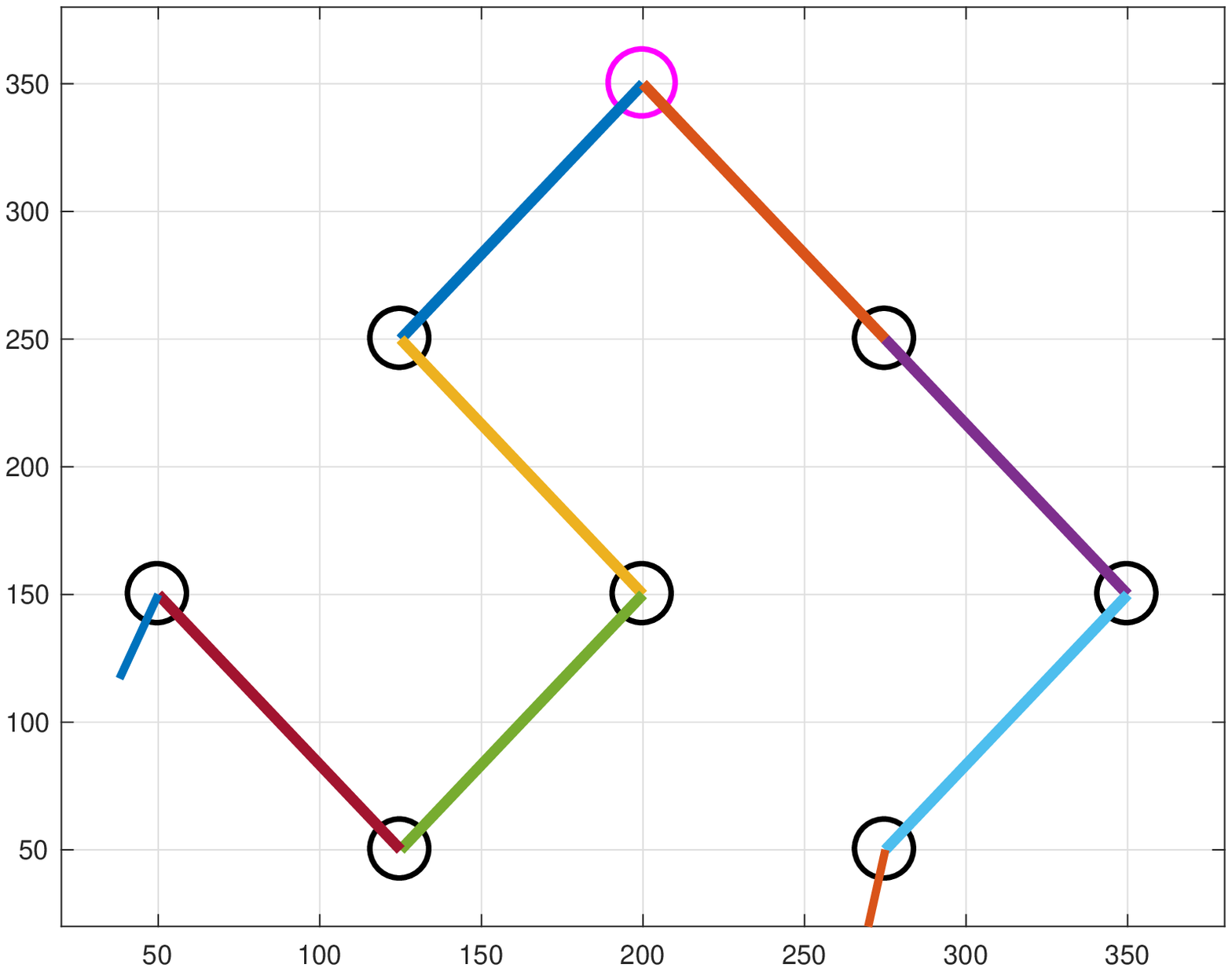}%
\caption{Single backup link ($k=3$)}%
\label{subfig-simple2}%
\end{subfigure}\hfill%
\begin{subfigure}{.5\columnwidth}
\includegraphics[width=\columnwidth]{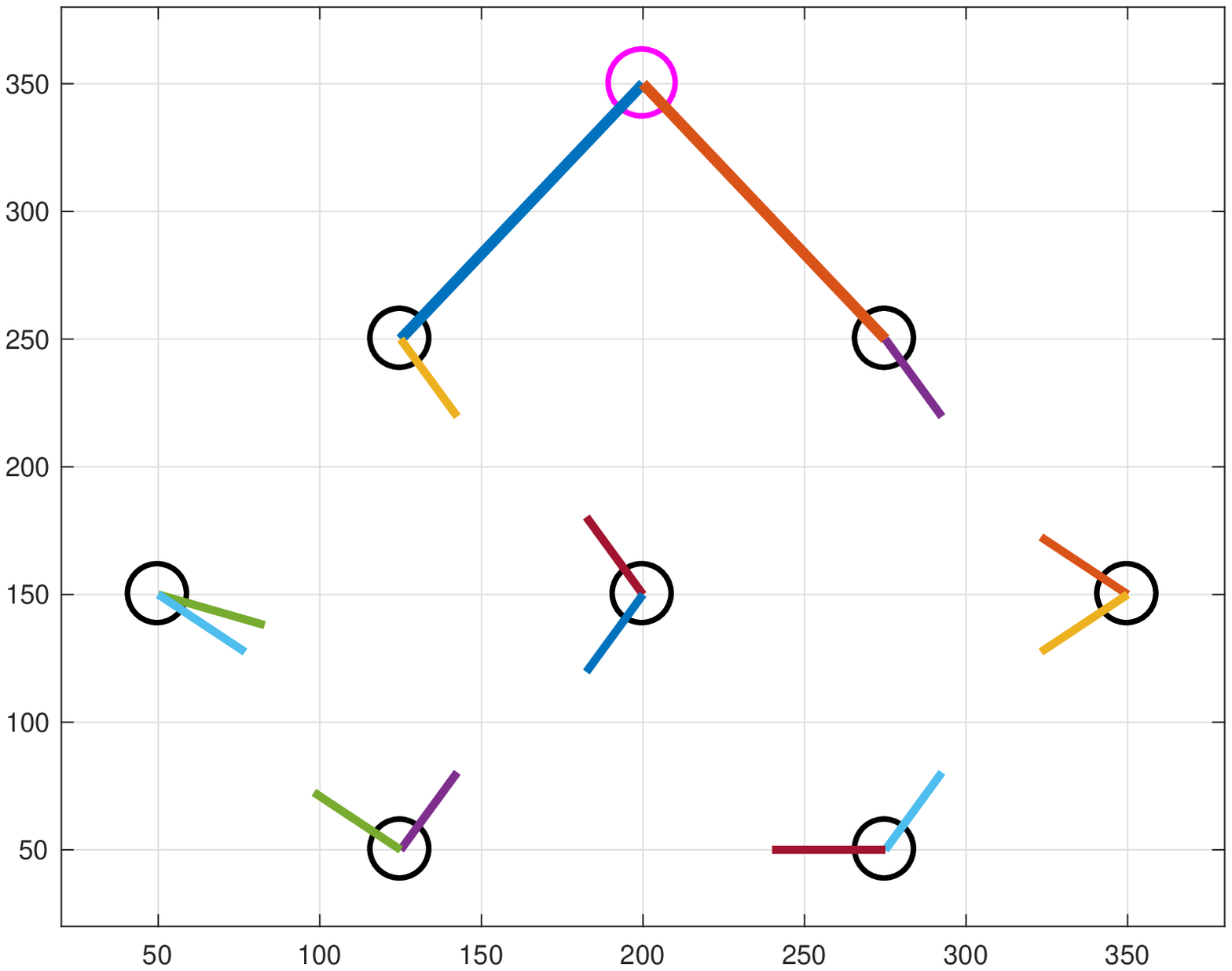}%
\caption{Major disruption ($k=13$)}%
\label{subfig-simple3}%
\end{subfigure}%
\begin{subfigure}{.5\columnwidth}
\includegraphics[width=\columnwidth]{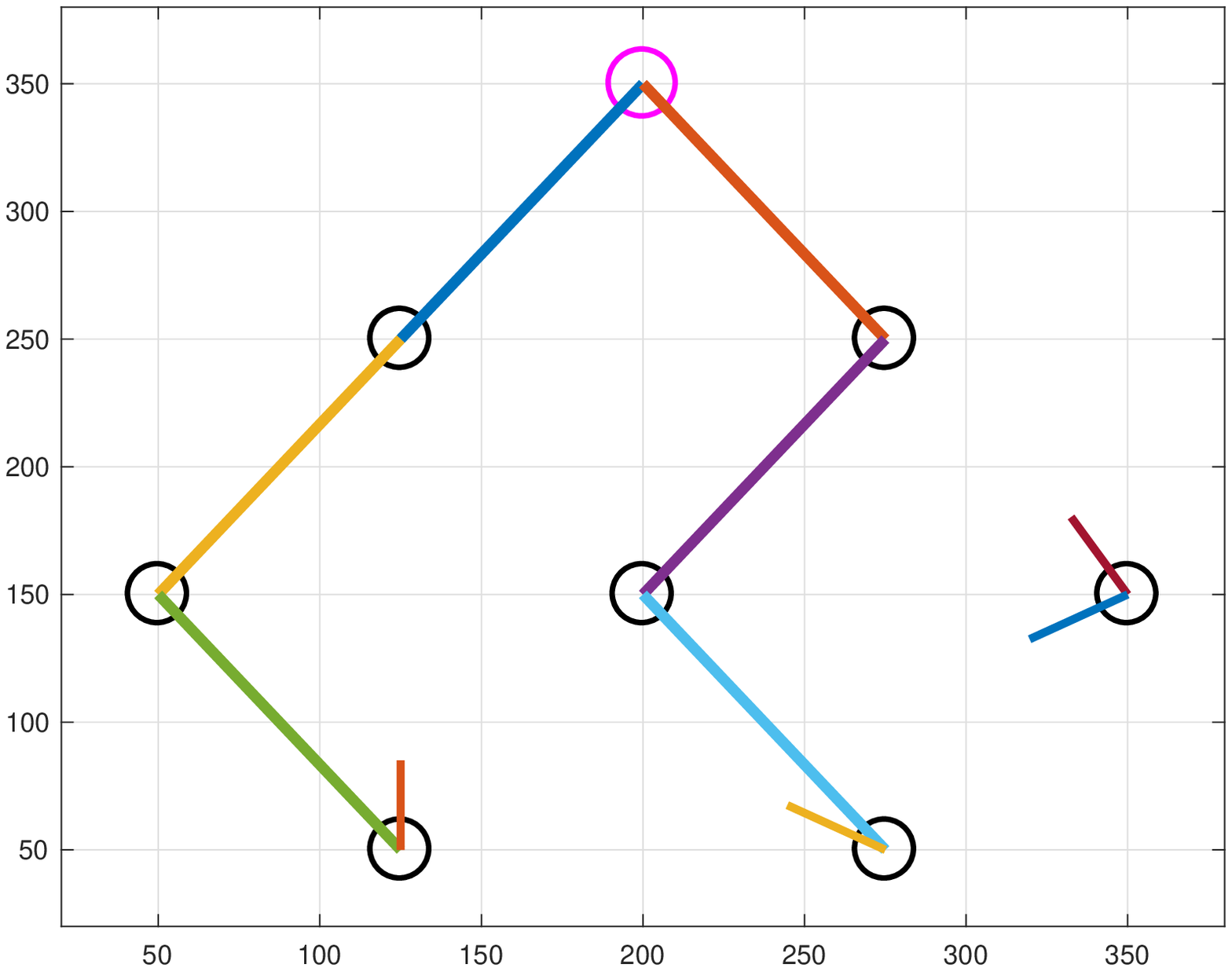}%
\caption{Final state ($k=K$)}%
\label{subfig-simple4}%
\end{subfigure}%
\caption{\textit{Simple} topology ($N=2$, $K=21$).}
\label{fig:simple-topology}
\end{figure*}

\begin{figure*}%
\centering
\begin{subfigure}{.5\columnwidth}
\includegraphics[width=\columnwidth]{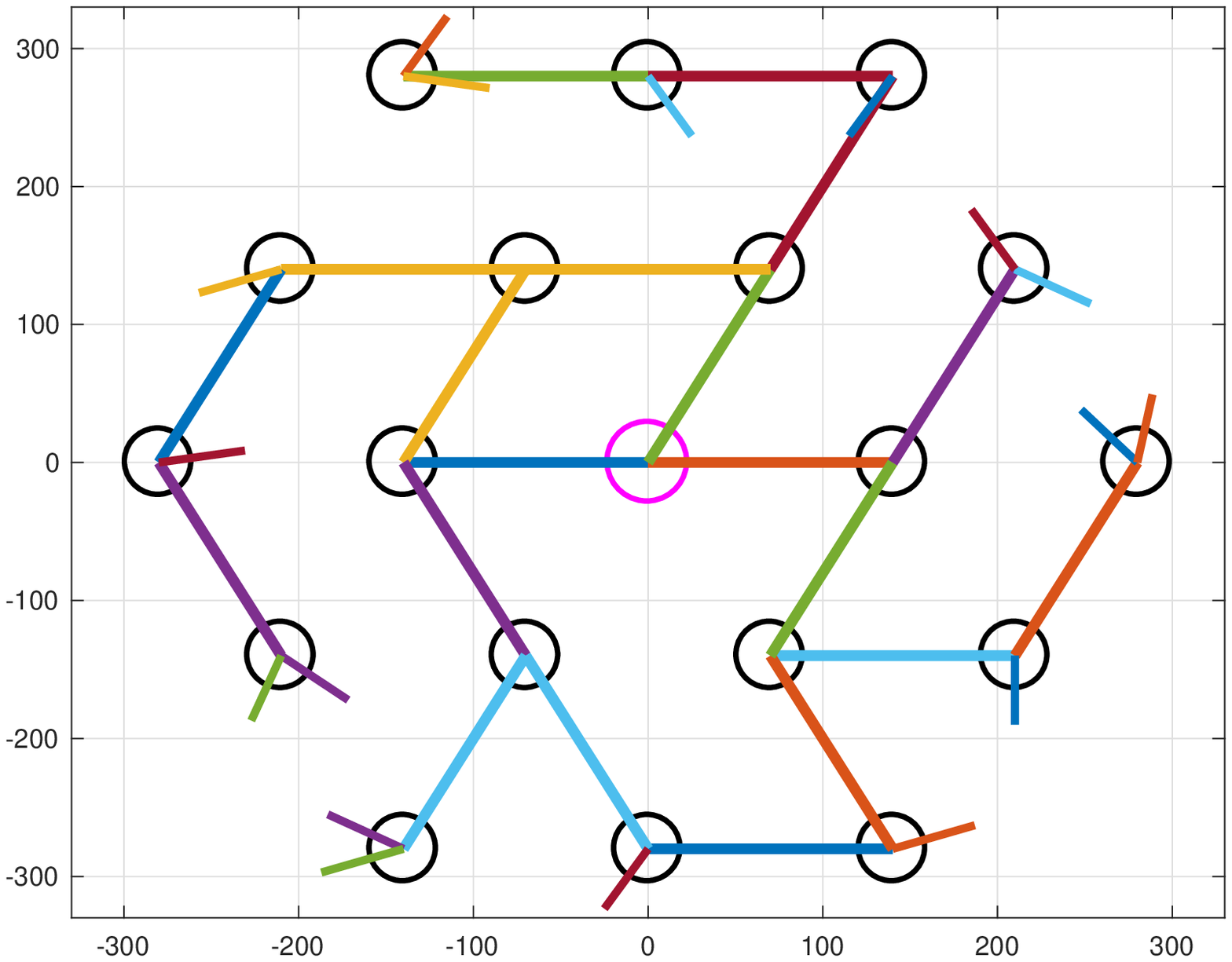}%
\caption{Initial state ($k=1$)}%
\label{subfig-hex1}%
\end{subfigure}\hfill%
\begin{subfigure}{.5\columnwidth}
\includegraphics[width=\columnwidth]{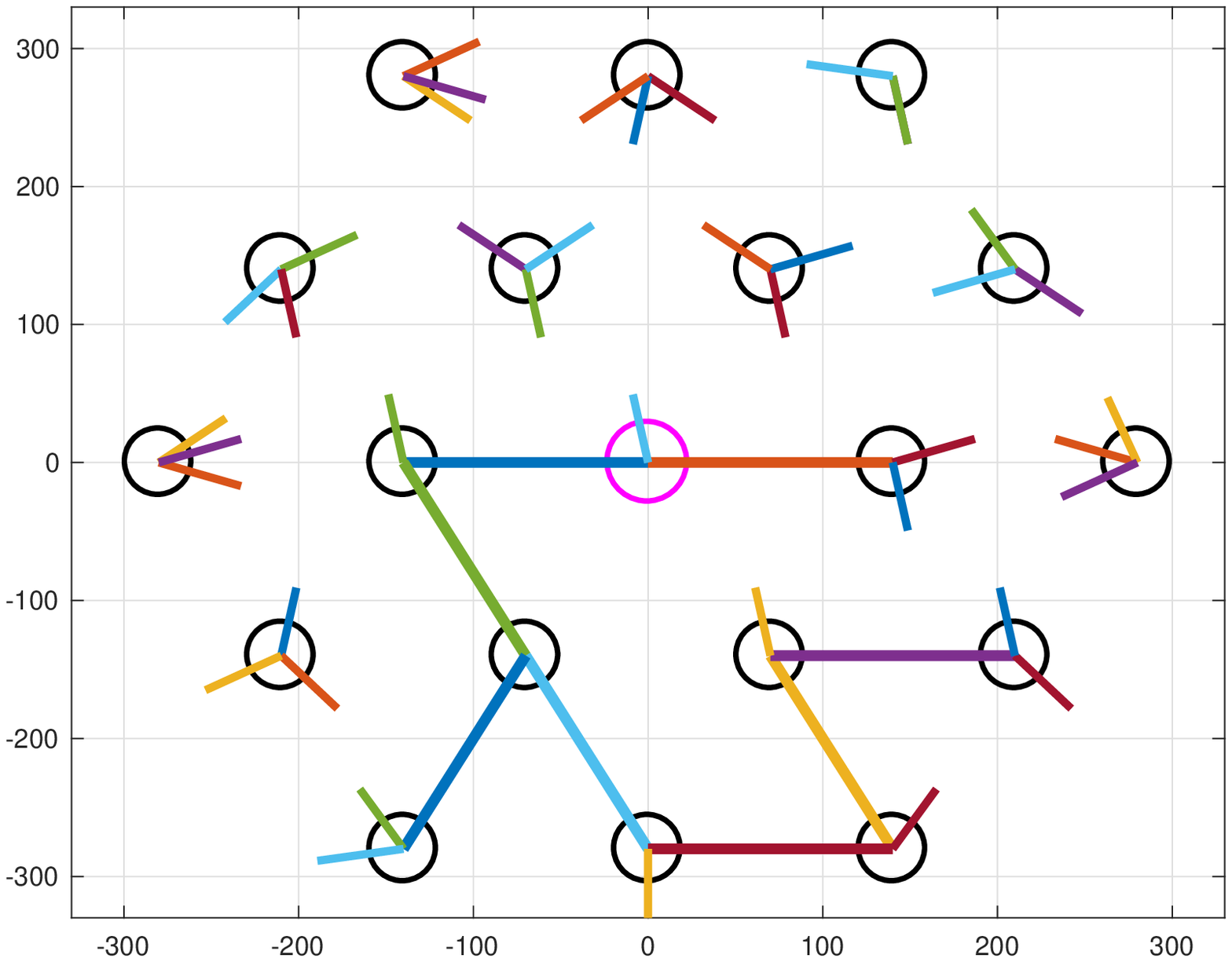}%
\caption{Major disruption ($k=5$)}%
\label{subfig-hex2}%
\end{subfigure}\hfill%
\begin{subfigure}{.5\columnwidth}
\includegraphics[width=\columnwidth]{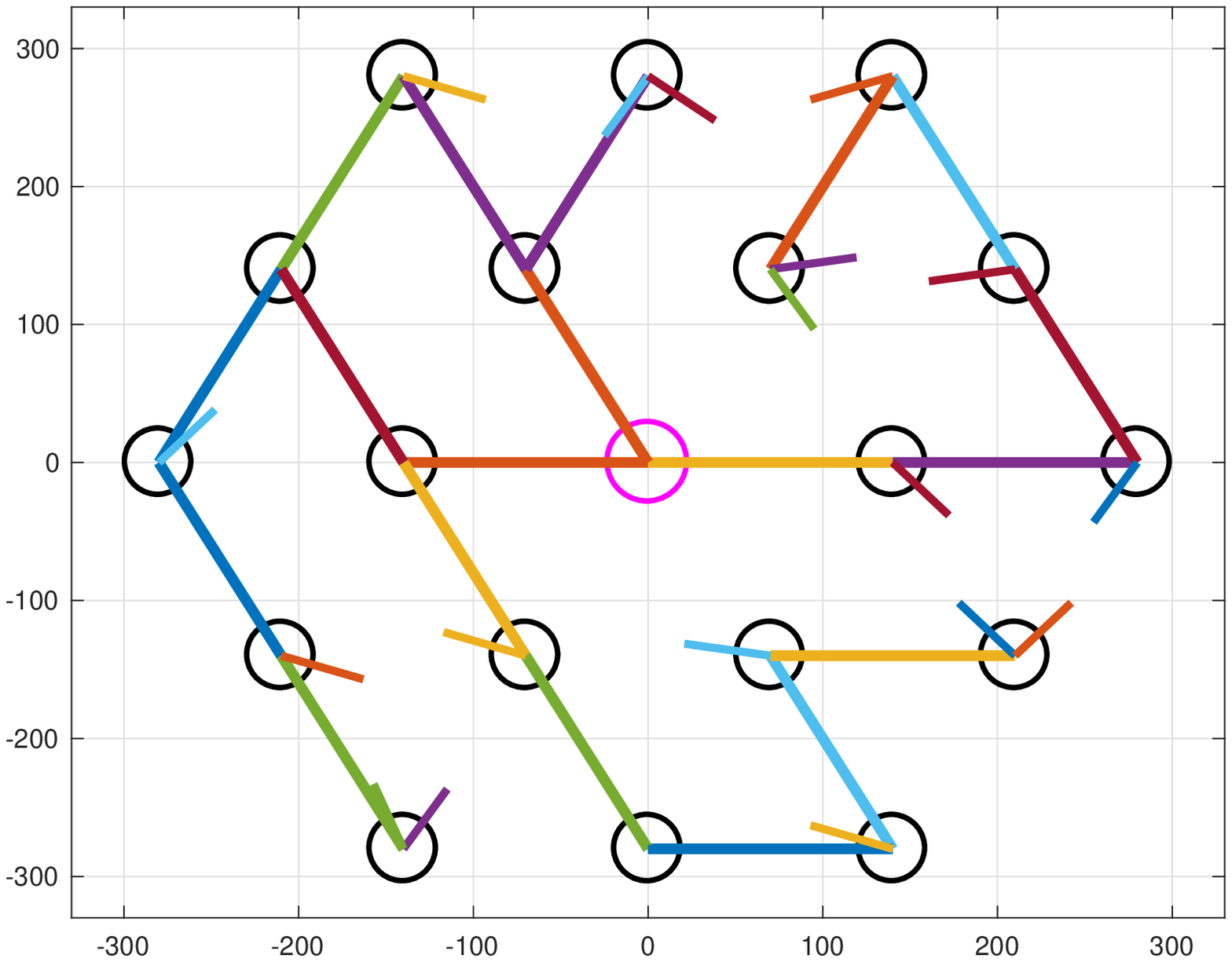}%
\caption{Backup path config. ($k=15$)}%
\label{subfig-hex3}%
\end{subfigure}%
\begin{subfigure}{.5\columnwidth}
\includegraphics[width=\columnwidth]{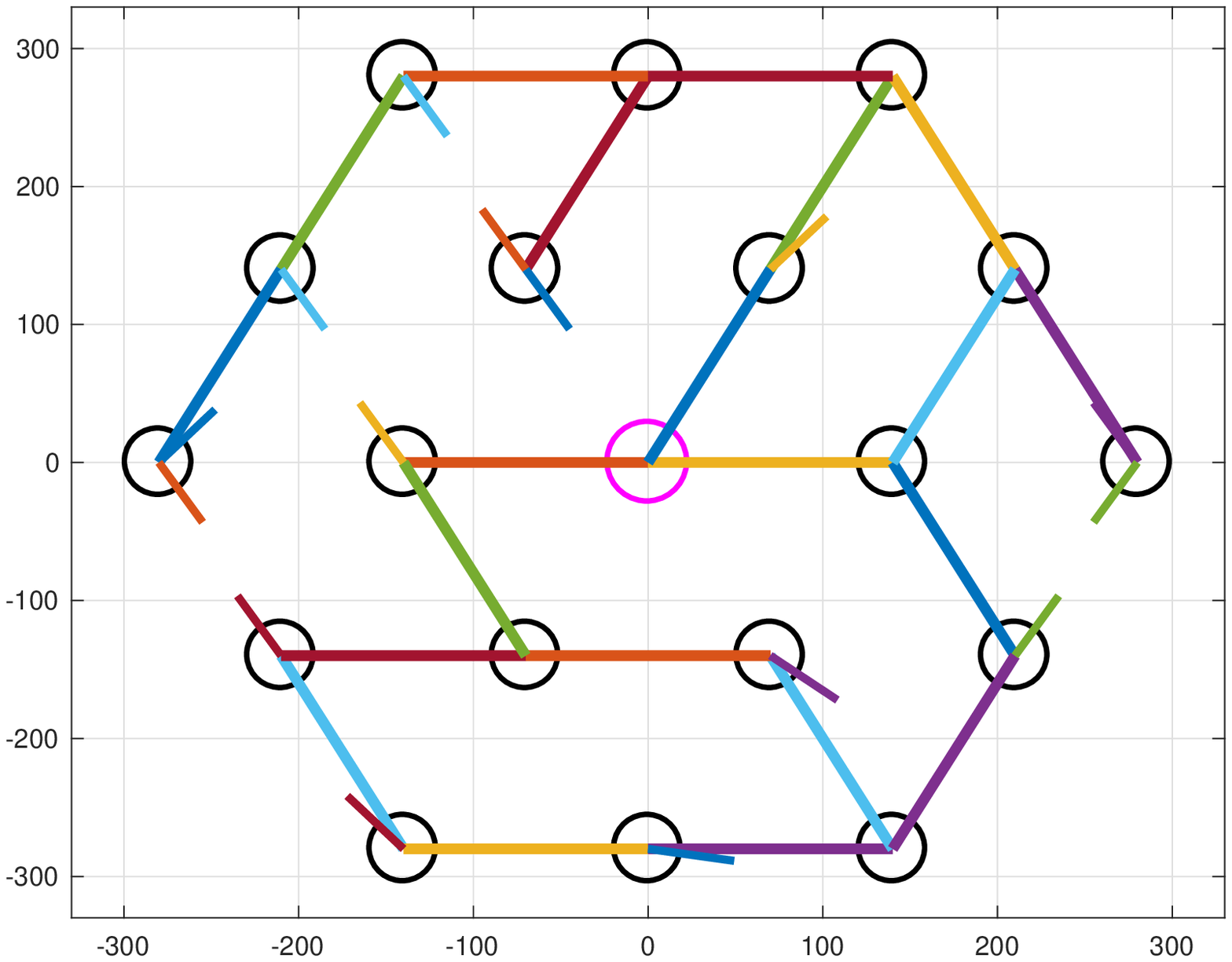}%
\caption{Final state ($k=K$)}%
\label{subfig-hex4}%
\end{subfigure}%
\caption{\textit{Hexagonal} topology ($N=3$, $K=21$).}
\label{fig:hex-topology}
\vspace{-0.2cm}
\end{figure*}

The \textit{Simple} topology (Fig.~\ref{fig:simple-topology}) was designed for an easy visualization of the operation of the framework, with 8 nodes (each with 2 interfaces) and 2 different 3-hop paths from the core-connected node (at the top of the topology) to the 2 mesh nodes at the bottom of the network. Every node has the closest next-hops located within 125 m. The initial configuration has one unused mesh node (Fig.~\ref{subfig-simple1}) that is used in the final configuration (Fig.~\ref{subfig-simple4}), forcing the model to adapt the network to this change. The \textit{Grid} topology features a $4\times4$ mesh where every node is placed over every $s$ =  180 m and then shifted on the $x$ and $y$ axis by two independent random variables following a normal distribution, with $\mu=0$ and $\sigma=\frac{s}{8}$. Finally, the \textit{Hexagon} topology (Figure~\ref{fig:hex-topology}) follows a hexagonal layout with 19 nodes \cite{CBR}, spaced with 140 m increments on both $x$ and $y$ axis.

For simplicity and clarity purposes, we use $\theta=10$ and select the minimum $K=19$ by considering the worst case scenario on every topology, i.e. when an antenna needs to perform a 180\textdegree ~rotation. We then select $K=20$ and $K=21$, to assess how the results vary with additional available reconfiguration time slots.
Because the MILP is known to be NP-hard, larger $K$ values were not used, due to the excessive optimization solving time.
We select $\tau=0.2$ s, corresponding to a full 360\textdegree ~rotation in 7.2 s. As a follow-up, we can modify this value based on testbed measurements, with prototypes of mechanical platforms. The maximum capacity of each mmWave link is computed using the propagation and atmospheric path loss models given in \cite{mesodiakaki2016energy} for the 60 GHz band with a transmit power level of 23 dBm. A truncated Shannon equation is employed, limiting the data rate between 4.64 Gbps and 1 Gbps over a mmWave link, according to the channel's quality.

To create the traffic demands for each topology, we fix the number of users and assign them different demands, i.e. 70\% of the users require 50 Mbps, 20\% need 75 Mbps and 10\% have a 100 Mbps demand, respectively. We then assign every user demand to a random node $d \in \mathcal{D}$. The number of users and the sum of all the demands $\sum \rho_d$ were selected to be large enough to congest the links from node $d \in \mathcal{I}$ when $N=3$, to overload the links when $N=2$, and to provide a lower load when $N=4$. With exception of the \textit{Simple} topology, where the first and last configurations were fixed, the initial $\mathcal{X}_{init}$ and final $\mathcal{X}_{end}$ snapshots were generated by creating two different traffic demands with the previously described method, then computing the optimal network configuration for each of them (we calculate the $x^k_{dnd'n'}$ values with the proposed model without the \eqref{eq:movement}, \eqref{eq:init-x}, \eqref{eq:end-x} and \eqref{eq:init-angle} constraints, and with $K=1$). The $\mathcal{X}_{end}$ demand values were used to set $\rho_d, \forall d \in \mathcal{D}$ in the experiments. To populate $\mathcal{A}^0$ with the alignments for the initially unused interfaces, we randomly select values $[0, 360[$ (multiples of $\theta$).

\begin{table}
\begin{center}
\caption{Parametrization of the used topologies}
\label{table:topologies}	
\begin{tabular}{ c | c | c | c | c }
\textbf{Topology} & \textbf{$D$} & \textbf{$N$} & \textbf{Users} & \textbf{$\sum \rho_d$}  \\
\hline
\textit{Simple}						& 8									& 2  & 80	&	5000 Mbps	\\ \hline
\multirow{3}{*}{\textit{Grid}}				&  \multirow{3}{*}{16}	& 2  & \multirow{3}{*}{100}	&	\multirow{3}{*}{6400 Mbps}	\\
							 					& 										& 3	&										&													\\
							 					& 										& 4	&										&													\\ \hline
\multirow{3}{*}{\textit{Hexagon}}		&  \multirow{3}{*}{19}	& 2	& \multirow{3}{*}{105}	& 	\multirow{3}{*}{6650 Mbps}	\\
							 					& 										& 3	&										&													\\
							 					& 										& 4	&										&													\\ \hline
\end{tabular}
\end{center}
\vspace{-0.2cm}
\end{table}

We use three weight functions $m_k = f(k)$ in \eqref{eq:objective}, namely:
\begin{itemize}
\item \textbf{Constant:}  $f(k) = 1$,
\item \textbf{Linear:} $f(k) = 2k$,
\item \textbf{Exponential:} $f(k) = e^k$.
\end{itemize}

By adding weights that depend on the number of elapsed time slots, different reconfiguration strategies can be selected. For example, when $m_k$ converges to high values, the model is forced to reach the final state within less slots, although it may not establish backup paths, i.e. most of the interfaces will initially rotate at the same time and no links will be formed (Figure~\ref{subfig-hex2}). When not adding any weight ($m_k = 1$), the model will equally use all available $K$ time slots to reach the final configuration, resulting in lower overall loss values. In addition, to test a scenario where high loss rates are not permitted during some parts of the reconfiguration, we performed tests using the constraints from \eqref{eq:loss-constraint} with $m_k = 1$, by limiting the total loss to $50\%$ of the total demand $\sum \rho_d$ over the second half of the execution ($[K/2 \dots K]$).

\begin{figure}
\centering \includegraphics[width=1.0\columnwidth]{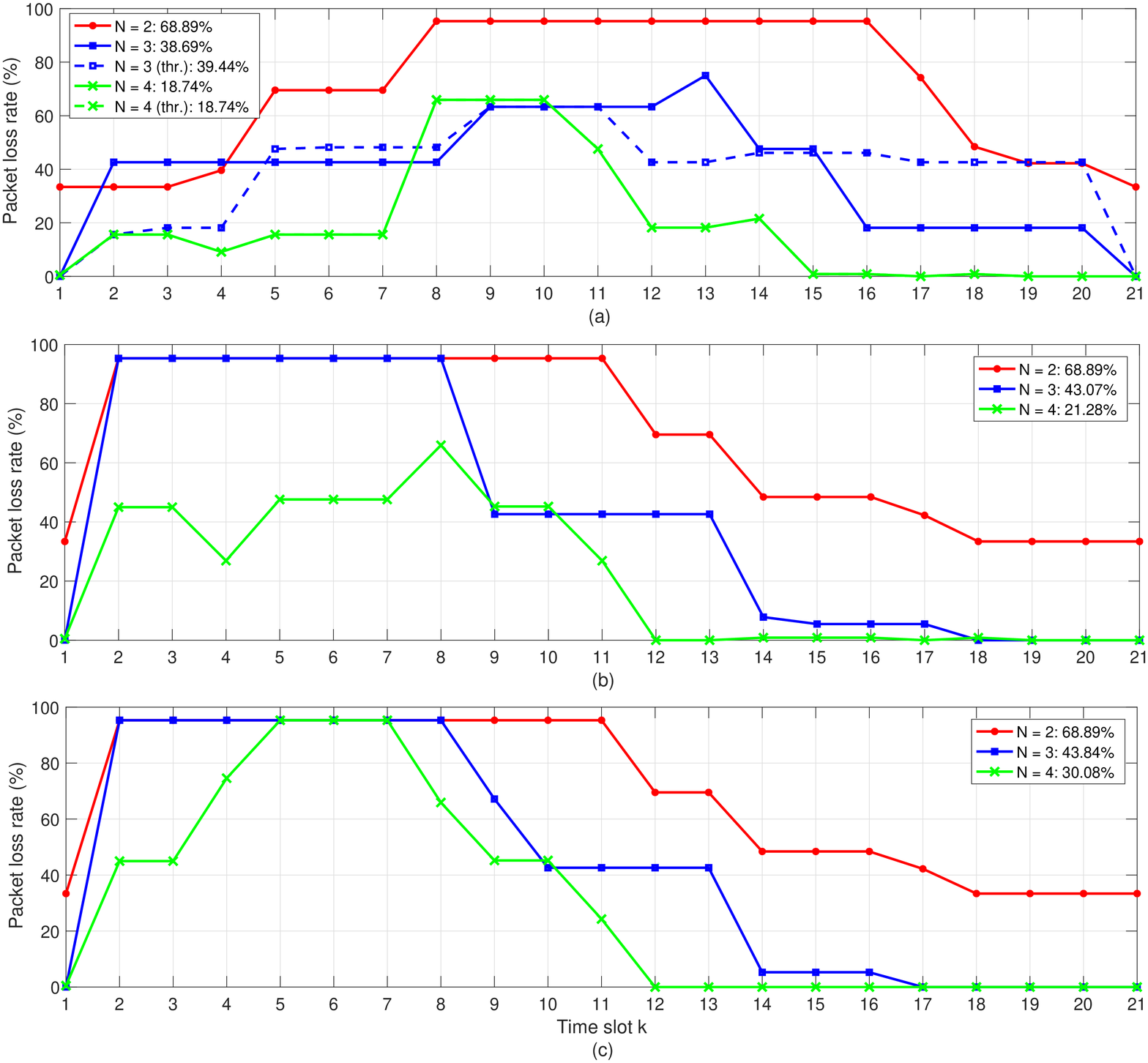}
\caption{Loss rate per time slot for the \textit{Grid} topology ($K=21$), where (a) $m_k=1$, (b) $m_k=2k$, and (c) $m_k=e^k$. In the legend, we indicate the loss rate for the total transition time.}
\label{fig:loss-over-time}
\vspace{-0.2cm}
\end{figure}

The impact of the different functions $m_k$ and loss thresholds with the \textit{Grid} topology and $K=21$ is depicted in Fig.~\ref{fig:loss-over-time}. When $N=2$, the network is overloaded and the average loss is close to 70\% with all the weight functions, often reaching over 95\% loss, when all the links from node $d \in \mathcal{I}$ are disabled. Therefore, it was infeasible to apply the 50\% loss threshold, as no solutions respecting this constraint could be calculated. With $N=3$, the loss rates decrease and the different behaviors of the weight functions are more pronounced. With $m_k = 1$, the topology is adjusted until $k=21$ and the loss rate smoothly varies over time. For the remaining weight functions, more than 95\% loss is observed in the early reconfiguration slots, but the topology converges to a 0\% loss state before $k=21$ ($k=18$ when $m_k = 2k$ and $k=17$ with $m_k = e^k$, respectively). With the loss thresholds and $m_k = 1$, no more than 50\% of loss occurs in the second half of the configuration, and the average loss rate difference was not significant, having nearly 1\% more loss than without any thresholds. Finally, with $N=4$, the same behavior as $N=3$ is observed, but with less overall loss, due to the higher link availability. Yet, due to the aggressive values of $e^k$, more than 95\% loss with $k=5,6,7$ can be observed, as latter loss occurrences would greatly increase the objective function value. For this scenario, applying the thresholds did not produce any different results with $m_k = 1$, since the threshold constraints were never violated.

\begin{figure}
\centering \includegraphics[width=1.0\columnwidth]{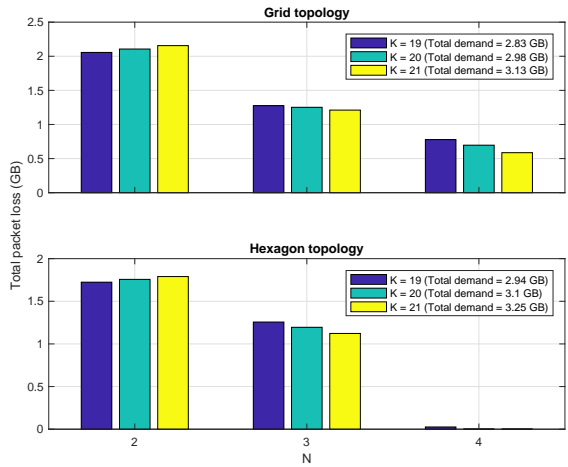}
\caption{Total loss versus the number of interfaces $N$ for different transition time periods $K$.}
\label{fig:total-loss}
\vspace{-0.2cm}
\end{figure}

Fig.~\ref{fig:total-loss} presents the total loss, $\tau \left[\sum_k^K \sum_d^D l^k_d \right] $ (GB), for $m_k = 1$, while varying the number of network interfaces $N$ and total reconfiguration time slots $K$.
For both \textit{Grid} and \textit{Hexagon} topologies, when $N=2$, the node $d \in \mathcal{I}$ cannot provide the total requested bandwidth, and consequently, it is not possible to reach a lossless configuration. With that, the total loss always increases when a larger $K$ is selected. With $N = 3$, it is possible to reach a final configuration with no loss at $k=K$, although it is not always possible to provide backup paths, as most of the interfaces always need to be used, provoking major disruption events when they need to rotate (e.g. Figure~\ref{subfig-hex2}). However, as we allow more time slots for reconfiguration, we can create intermediate backup paths and, instead of directly converging into the final state, the network can use temporary links while the interfaces from the final configuration rotate, contributing to less global network disruptions. Finally, when $N=4$, more interfaces can provide protection for links going down due to the rotation. Similarly to $N = 3$, with larger $K$ values it is possible to achieve more intermediate steps that use backup paths. As can be seen, there is still significant loss in the \textit{Grid} topology (0.78 GB, 0.70 GB and 0.59 GB for $K = 19, 20, 21$, correspondingly), since it is not possible to transition to the last configuration state without suffering high loss rates, as a high number of interfaces need to rotate during most of the time slots. For $N = 4$ with the \textit{Hexagonal} topology, the total loss is almost zero (27 MB with $K = 19$ and 1 MB with $K = 20$ and $21$), as it is possible to establish backup links to deliver the necessary traffic during most of the time slots. Specifically, with $K = 19$ the loss is higher since it is not possible to provide the total demand on the existing links during the initial time slots, while with $K = 20$ and $21$, the 1 MB lost is caused by the insufficient bandwidth with $X_{init}$ at $k=1$. In conclusion, $K$ should be large enough to provide auxiliary links during the configuration period, while the total transition time (according to $\tau$) should also be considered by the network operator.
Finally, as an extension of this work, we emphasis the importance of using larger $K$ values, when combining our framework with fast solution heuristics. This will allow finding the $K$ values that can produce the minimum achievable loss results, for a given topology.

\section{Conclusion}
\label{sec:conclusions}
In this paper, we developed a mathematical model that orchestrates the reconfiguration of a mesh based SC BH network, composed of multi-hop mmWave links that are controlled by rotational mechanical platforms, which need to be aligned with each other to form links. The model calculates the optimal intermediate reconfiguration steps necessary to transition between two configuration snapshots, while minimizing the total packet loss. The evaluation of the developed framework with different topologies, number of nodes, number of interfaces per node, and different configuration times shows that increasing the number of interfaces allows the computation of backup paths, which decrease the total loss rate. Moreover, when increasing the allowed reconfiguration time, the proposed framework also diminishes the overall loss, as the additional time allows the interfaces to rotate and form more intermediate protection links. Due to the solved problem's complexity, we will develop fast solution heuristics, suitable for online optimization. In addition, we will perform experimental work on a mmWave mesh testbed, where our steerable mmWave mesh BH reconfiguration approach can be triggered by an SDN controller.

\section*{Acknowledgement}
Parts of this work have been funded by the Knowledge Foundation of Sweden through the project SOCRA.

{\small{}\bibliographystyle{/IEEEtran}
\bibliography{globecom2018bib}
}{\small \par}


\end{document}